\documentclass[10pt,journal,compsoc]{IEEEtran}
\IEEEoverridecommandlockouts
\usepackage{cite}
\usepackage{amsmath,amssymb,amsfonts}
\usepackage{algorithmic}
\usepackage{graphicx}
\usepackage{textcomp}
\usepackage{multirow}
\usepackage{xcolor}
\usepackage{booktabs}
\usepackage{array}
\usepackage{float}
\usepackage{hyperref}
\usepackage{verbatim}
\usepackage{bm}
\usepackage{listings}

  \usepackage{cite}

\newcommand{\tabincell}[2]{\begin{tabular}{@{}#1@{}}#2\end{tabular}}
\def\BibTeX{{\rm B\kern-.05em{\sc i\kern-.025em b}\kern-.08em
    T\kern-.1667em\lower.7ex\hbox{E}\kern-.125emX}}

\newcommand{\ematrix}{\mathcal{M}}

\begin{document}

\title{Mutation Testing for Ethereum Smart Contract}

\author{Haoran~Wu, Xingya~Wang, Jiehui~Xu, Weiqin~Zou, Lingming~Zhang, and~Zhenyu~Chen \IEEEcompsocitemizethanks{\IEEEcompsocthanksitem  Haoran~Wu, Xingya~Wang, Jiehui~Xu, Weiqin~Zou, and~Zhenyu~Chen are with the State Key Laboratory for Novel Software Technology, Nanjing University, 210093, China. E-mail: \{xingyawang, zychen\}@nju.edu.cn
\IEEEcompsocthanksitem Lingming~Zhang is with the Department of Computer Science, University of Texas at Dallas, USA.}
}

\maketitle

\begin{abstract}
Smart contract is a special program that manages digital assets on blockchain. It is difficult to recover the loss if users make transactions through buggy smart contracts, which cannot be directly fixed. Hence, it is important to ensure the correctness of smart contracts before deploying them.
This paper proposes a systematic framework to mutation testing for smart contracts on Ethereum, which is currently the most popular open blockchain for deploying and running smart contracts. Fifteen novel mutation operators have been designed for Ethereum Smart Contracts (ESC), in terms of keyword, global variable/function, variable unit, and error handling. An empirical study on 26 smart contracts in four Ethereum DApps has been conducted to evaluate the effectiveness of mutation testing. The experimental results show that our approach can outperform the coverage-based approach on defect detection rate (96.01\% vs. 55.68\%). The ESC mutation operators are effective to reveal real defects and we found 117 out of 729 real bug reports are related to our operators. These show the great potential of using mutation testing for quality assurance of ESC.
\end{abstract}


\section{Introduction}
Blockchain, the foundation of Bitcoin, has been gaining increasing attention from both industry and research community \cite{zheng2018blockchain}.
As an immutable ledger, the blockchain technology allows transactions to take place in a decentralized manner \cite{nakamoto2008bitcoin}.
Many blockchain-based applications beyond cryptocurrencies like Bitcoin have been proposed.
An emerging area of blockchain technology is smart contract \cite{luu2016making}.
A smart contract is a special program (running on blockchains) that controls users' digital assets.
It could be automatically and correctly executed by a network of mutually distrusting nodes without the need of an external trusted authority \cite{zheng2018blockchain}.
Smart contract has increasingly been finding numerous important applications (e.g., crowdfunding, voting, gaming, etc) in the real world \cite{bartoletti2017empirical}. More and more developers are devoting themselves to developing various kinds of smart contracts. For example, there have been as many as 1 million smart contracts being deployed and running on Ethereum -- the most popular open blockchain platform for deploying and running smart contracts \cite{Wang2019}.

Despite the increasing popularity of smart contracts applications in practice, the quality of smart contracts was reported to be worrisome. According to Nikoli et al. study, among the sampled 3,759 smart contracts, 3,686 smart contracts had an 89\% probability of containing vulnerabilities \cite{nikolic2018}.
Any potential vulnerability within a smart contract may lead to a significant financial loss.
For example, hackers stole as much as \$150 million from Ethereum by exploiting a loophole within the DAO contract \cite{Ammous2016}.
On the other hand, due to the nature of blockchain, any transactions through smart contacts could not be reverted. Which means if users experience loss on blockchain, they cannot recover their losses.
In addition, the code of smart contract is immutable once deployed onto blockchain.
All these called for effective approaches to guard/evaluate the quality of smart contracts in practice.


Compared with conventional software systems, the environment for developing ESC is still immature, 16 and most developers are young and lack of experience. Nikoli et al. sampled 3,759 smart contracts, of which 3,686 smart contracts had a 89\% probability of containing vulnerabilities \cite{nikolic2018}, indicating that developing a high-quality ESC is full of challenges. As stated, a small defect might result in extremely serious consequences. Meanwhile, testing is widely recognized as one of the most important means of ensuring software quality \cite{Orso2014}. Thus, the Smart Contract Under Deployment (SCUD) must be given adequate testing, regardless of cost, aiming to detect defects as more as possible. Subsequently, the following question should be answered, that is, how to evaluate the adequacy of ESC test suites? Mutation testing provides a costly but effective approach for test adequacy evaluation~\cite{delgado2017assessment}. It achieves this by checking the ability of the given tests to reveal some artificial defects \cite{papadakis2019mutation}. Currently, mutation testing has been applied on a series of programming languages such as C \cite{Chekam2017}, C++ \cite{delgado2017}, Java \cite{pitest2014} and JavaScript \cite{Mirshokraie2013}, and has also been adapted for a set of programming paradigms such as Object-Oriented \cite{ma2002}, Functional \cite{le2014} and Aspect-Oriented \cite{omar2012} programming. Therefore, it is desirable to employ mutation testing for evaluating ESC test adequacy. 


Testing approaches were often used to guard/evaluate the quality of software programs.
As an effective testing approach, mutation testing was commonly used within traditional software programs \cite{Chekam2017, delgado2017, pitest2014}.
However, the potential of mutation testing has not been explored in the context of smart contracts.
Towards this end, we proposed to apply mutation testing on Ethereum Smart Contract (ESC).
Specifically, we designed a systematical framework for mutating Ethereum smart contract.
Our framework included four parts. First, for a given Smart Contract Under Test (SCUT), we firstly transformed it into an Abstract Syntax Tree (AST) and performed mutation on the AST.
Then, the mutated ASTs were transformed back to their corresponding source code version mutants.
Next, we built a testnet for each mutant, during which the testnets was initialized with the same state and contracts that SCUT depended on or relied on SCUT were deployed in a specific order. 
After that, we executed each mutant, recorded their execution results on the given test suite, and collected the mutation scores as the final evaluation results.

An important part of mutation testing is to define mutation operators, as mutation operators defined how to introduce syntactic changes to the original programs \cite{papadakis2019mutation}. 
The goal of mutation operators is to simulate the potential threats as completely as possible, and different operators are generally needed if different programming languages are adopted to develop programs \cite{papadakis2019mutation}.
Unlike traditional programs that were developed in general-purpose programming languages such as Java and Python, 
smart contracts were developed in specific-purpose languages such as Solidity or Vyper \cite{Smart-Contract-Languages}.
To adapt to new programming languages of smart contracts, we designed 15 novel mutation operators (in addition to 10 existing mutation operators) in terms of keyword, global variable/function, variable unit, and error handling, to effectively mutate smart contracts.

Twenty-six smart contracts from four Ethereum Decentralized Applications (DApps) were used to evaluate the effectiveness of our proposed mutation testing framework.
We found that our approach could outperform the coverage-based approach on defect detection rate (96.01\% vs. 55.68\%).
And the newly-proposed 15 mutation operators were effective to reveal real defects, where 117 real bug reports were related to our 15 mutation operators.
These results to a great extent showed the great potential of using mutation testing to guard the quality of Ethereum smart contract.

Our main contributions are as follows:

\begin{itemize}
\item\textbf{Idea.} We introduce the idea of applying mutation testing to ensure the quality of ESC. 
\item\textbf{Implementation.} We implement the first ESC-oriented mutation testing system, and design 15 novel mutation operators to simulate the potential defects in ESC. 
\item\textbf{Study.} We carried out an empirical study on four real-world DApps. The study results verified that mutation testing could indeed help improve ESC testing evaluation, and the novel operators were able to simulate real ESC defects well.
\end{itemize}

The rest of this study is organized as follows. Section II describes the basics of Ethereum smart contract and mutation testing. Our mutation testing approach is presented in detail in Section III, and each novel mutation operator is detailed in Section IV. Experimental design and results analysis are presented in Section V. After that, we discuss the threats to validity and give the related work in Section VI and VII, respectively. Finally, Section VIII concludes the paper and outlines directions for future research.

\section{Background}

\subsection{Ethereum Smart Contract}

Smart contract encapsulates predefined state and transition rules, scenarios that trigger contract execution (e.g. arrival at a specific time or occurrence of a specific event), and response actions under a specific scenario. It was first proposed by Nick Szabo, who aims to flexibly create and manage intellectual assets in a decentralized environment \cite{Szabo1996}. Blockchain provides the first feasible decentralized technique, making it possible to automatically and correctly execute smart contracts without relying on a trusted authority \cite{zheng2018blockchain}. Currently, many mainstream blockchain systems (e.g., BitCoin and Ethereum) have supported smart contract technique. With its Turing-complete programming languages and a series of sophisticated building tools, Ethereum has become a reasonably attractive blockchain platform for constructing various kinds of Decentralized Applications (Dapp). 

\begin{figure}[!htbp]%
	\begin{center}
	\includegraphics[width=0.95 \linewidth]{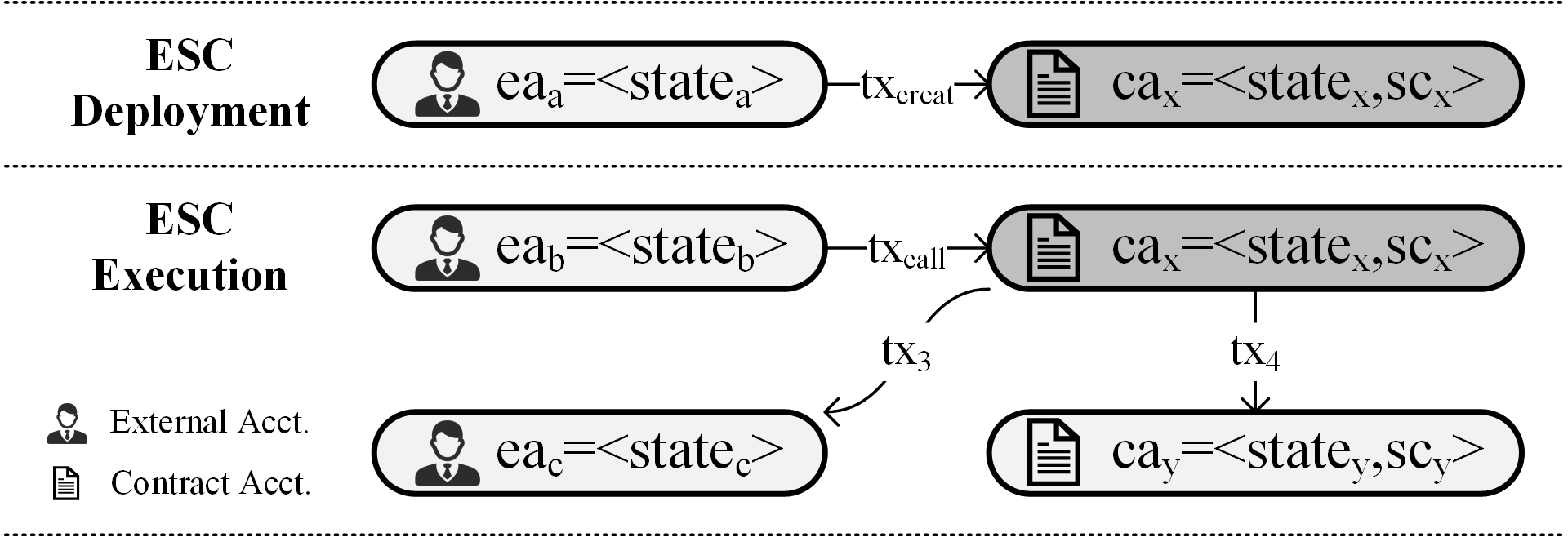}
	\caption{ESC Deployment and Execution}
	\label{ESC-Deployment-and-Execution}
	\end{center}
\end{figure}

Ethereum users rely on accounts to carry out transactions. There are two types of accounts, External Account (EA) and Contract Account (CA), in Ethereum. The former is controlled by the users with private keys, and the latter is controlled by ESC code \cite{Buterin2014}. Both EA and CA record the state information such as account balance, and CA has an additional unchanged field, codeHash, which points to the corresponding ESC. CA creation and its corresponding ESC execution are transaction-driven. An Ethereum transaction can be denoted by a quadruple $tx=\langle from, to, value, data \rangle$, where $from$ and $to$ denote the addresses of sender's account and receiving account respectively, and $value$ denotes the number of Wei to be transferred to $to$. $tx$ is an CA creation transaction when $to=\varnothing$. At this time, $data$ denotes SCUD. Otherwise (i.e., $to\neq\varnothing$), if $to$ is a CA, $data$ denotes the input to $to$'s corresponding ESC code. ESC deployment can be deemed as the process of CA creation. Fig.\ref{ESC-Deployment-and-Execution} depicts the ways of deploying as well as executing the ESC $sc_x$. 

\begin{itemize}
\item[(1)] \textbf{ESC Deployment}. EA $ea_a$ launches a CA creation transaction $tx_{create}=\langle addr_a, \varnothing, value_1, sc_x \rangle$. Once $tx_{create}$ has been packed into a block and the block has been persisted in Ethereum, CA $ca_x$ creation as well as $sc_x$ deployment are finished. 
\item[(2)] \textbf{ESC Execution}. EA $ea_b$ launches a message call transaction $tx_{call}=\langle addr_b, addr_x, value_2, input_x \rangle$. Once $tx_{call}$ has been packed into a block and the block has been persisted in Ethereum, $sc_x$ takes $input_x$ as input and starts to run. Executing $sc_x$ results in two additional transactions, $tx_3$ and $tx_4$. 
\end{itemize}

Ethereum provides four smart contract programming languages (i.e., Solidity, Vyper, Bamboo and Flint), where each of them is influenced by one or several popular languages \cite{Smart-Contract-Languages}. For example, Solidity is mainly influenced by JavaScript. Among them, Solidity is the most popular language on Ethereum. Different from JavaScript, which is often used in developing Web applications and runs in the Web browser, Solidity is specific to ESC and runs in the Ethereum Virtual Machine (EVM). Thus, though JavaScript and Solidity share a similar syntax, they still have a number of differences, which are detailed as follows: 

\begin{table*}[!htbp]%
	\centering 
	\caption{Specific Variable Units in Solidity}
	\label{Specific-Variable-Units-in-Solidity}
	\begin{tabular*}{\hsize}{@{}@{\extracolsep{\fill}}p{2.1cm}|p{4.4cm}|lp{2.1cm}@{}}
		\toprule
		\textbf{Unit Type}	& \textbf{Suffix} 						& \textbf{Description} & \\
		\midrule
       Ether            		& wei, finney, szabo, ether 				& 1 szabo = 1e12 wei, 1 finney = 1e15 wei, 1 ether = 1e18 wei & \\
       Time  					& seconds, minutes, hours, days, weeks 	& 1 minutes = 60 seconds, 1 hours = 60 minutes, 1 days = 24 hours, 1 weeks = 7 days & \\
		\bottomrule 
	\end{tabular*}
\end{table*}

\begin{table}[!htbp]%
	\centering 
	\caption{Specific Keywords in Solidity}
	\label{Specific-Keywords-in-Solidity}
	\begin{tabular*}{\hsize}{@{}@{\extracolsep{\fill}}p{2.1cm}|p{1cm}|p{4.9cm}@{}}
		\toprule
		\textbf{Specifier Type}	& \textbf{Keyword} & \textbf{Description} \\
		\midrule
        \multirow{4}{*}{function visibility} & public 			 & accessed by all \\
       						      	         & external          & accessed by using this and externally \\
       						      	         & internal 		 & accessed by this and derived contracts \\
       						      	         & private  		 & accessed by only this contract \\
        \hline
        \multirow{3}{*}{function state}      & pure 			 & not view and modify state \\
       				             	         & view 			 & not modify state \\
       				             	         & payable   		 & receive ethers \\
        \hline
        \multirow{3}{*}{reference location}  & memory 			 & lifetime is limited to a function call \\
       						      	         & storage 			 & persisted into the blockchain itself \\
       								         & calldata			 & required for params of external functions \\
		\hline
        lvalue                               & delete			 & assigns the initial value for its type \\
		\bottomrule 
	\end{tabular*}
\end{table}

\begin{itemize}
\item[(1)] \textbf{Variable Type}. Different from JS, Solidity is a statically and strongly typed language, indicating that the variable's type is specified during compilation (executing for JavaScript) and implicit type conversion is not allowed. Moreover, for ease of handling blockchain data, some novel variable types (e.g., address) are added in Solidity.  
\item[(2)] \textbf{Variable Unit}. Ethereum employs an intrinsic currency, Ether, to incentive computation within the network \cite{wood2014ethereum}. Then, as Table \ref{Specific-Variable-Units-in-Solidity} depicts, it defines some sub-denominations of Ether. It also defines some time units. 
\item[(3)] \textbf{Keyword}. As Table \ref{Specific-Keywords-in-Solidity} depicts, Solidity defines a set of novel keywords in terms of function visibility, function state, reference location, and lvalue to optimize the storage and reduce the gas cost of ESC. 
\item[(4)] \textbf{Global Variable and Global Function}. For ease of accessing blockchain data (e.g., block number, difficulty and gasLimit) and carrying out mathematical calculations (e.g., \texttt{addmod}, \texttt{mulmod} and \texttt{keccak256}), Solidity provides a set of global variables and global functions. 
\item[(5)] \textbf{Error Handling}. In addition to assert functions, Solidity provides \texttt{require($condition$)} for handling errors. ESC continues when $condition$ is satisfied and otherwise stops. Function \texttt{require} is used for variable checking, while \texttt{assert} is mainly used to detect unknown errors.  
\end{itemize}

\subsection{Mutation Testing}

Mutation testing is a fault-based software testing technique which can be used to evaluate the adequacy of test cases. These so-called mutants are based on well-defined mutated operators that either simulate typical application errors or force a validity test. The goal is to help testers identify limitations in the testing process or test suite.

When applying mutation testing, testers first design mutation operators according to the characteristics of the program under test. Generally, mutation operators only make minor changes to the program under test in accordance with the grammar. Applying mutation operators to the program under test can generate a large number of defective programs called mutants. Then the \emph{equivalent} mutants (i.e., the mutants that have no effect on the execution result of the program) are excluded. Next, the original program and each mutant are tested with a given test suite. If the test result of a mutant is different from the original program, it is said to be \emph{killed}. This indicates that existing test cases can detect the defect. If the test result of a mutant is the same as that of the original program, it is said to be \emph{survived}, and new test cases need to be designed to kill it. Finally, mutation testing calculates the mutation score according to the following formula: $\fontsize{8pt}{8pt}\frac{TotalNonEquivMutants - SurvivingMutants}{TotalNonEquivMutants}\times 100\%$. The higher the mutation score, the higher the adequacy of test cases. 


Mutation testing has been shown to subsume other test criteria by incorporating appropriate mutation operators~\cite{walsh1985, frankl1997, andrews2005, just2014}. Thus, designing effective mutation operators is one of the most important tasks when applying mutation testing to a new field\cite{deng2017}. In this paper, we first summarize a set of general mutation operators that can be used in Solidity based on the existing mutation operators for JavaScript \cite{Mirshokraie2013}, and then propose several specific mutation operators according to the characteristics of Solidity.

\begin{table}[htbp]
	\centering 
	\caption{ General Mutation Operators}
	\label{General-mutation-operators}
	\begin{tabular}{c|l}
		\toprule
		\textbf{Operator} & \textbf{Description}\\
		\hline
		AOR    & Arithmetic Operator Replacement \\ \hline
		AOI    & Arithmetic Operator Insertion \\ \hline
		ROR    & Relational Operator Replacement \\ \hline
		COR    & Conditional Operator Replacement \\ \hline
		LOR    & Logical Operator Replacement \\ \hline
		ASR    & Assignment Operator Replacement \\ \hline
		SDL    & Statement Deletion \\ \hline
		RVR    & Return Value Replacement \\ \hline
		CSC    & Condition Statement Change \\
		\bottomrule
	\end{tabular}
\end{table} 

Table \ref{General-mutation-operators} shows the general mutation operators. AOR is divided into two kinds of mutation operators, AOR$_B$ (Binary Arithmetic Operator Replacement) and AOR$_S$ (Short-cut Arithmetic Operator Replacement). The AOR$_B$ operator replaces basic binary arithmetic operators ($+$, $-$, $*$, $/$, and $\%$) with other binary arithmetic operators while the AOR$_S$ operator replaces short-cut arithmetic operators (op$++$, $++$op, op$--$, and $--$op). The ROR operator replaces relational operators ($>$, $>=$, $<$, $<=$, $==$, and $!$$=$) with other relational operators or replace the entire predicate with \emph{true} and \emph{false}. The COR operator replaces binary conditional operators ($\&\&$, $||$, $\&$, $|$) with other binary conditional operators. The ASR operator replaces short-cut assignment operators ($+$$=$, $*$$=$, $/$$=$, $\%$$=$, and $\&$$=$) with other short-cut assignment operators. The SDL operator can delete an executable statement by commenting it out, and the CSC operator can force the statement in a conditional judgment to be \emph{true} or \emph{false}.

\section{Mutating Smart Contracts}

\begin{figure*}[!htbp]%
\begin{center}
\includegraphics[width=0.95\linewidth]{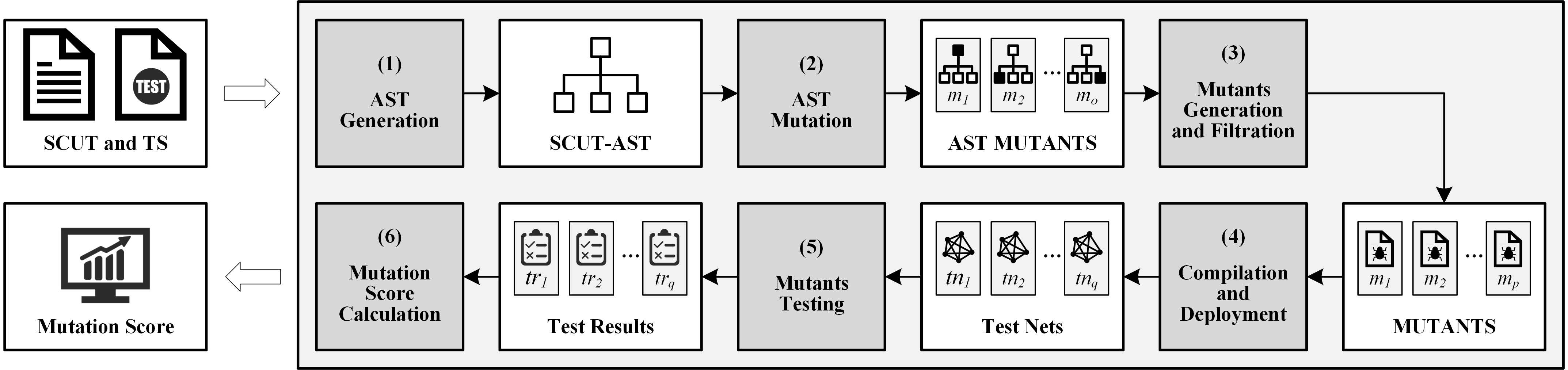}
\caption{Performing Mutation Analysis on Smart Contracts}
\label{Performing-mutation-analysis-on-smart-contracts}
\end{center}
\end{figure*}

Traditional JavaScript mutation engine usually parses the source file to an AST (Abstract Syntax Tree), and then performs mutant generation on it \cite{mirshokraie2015guided}. The generated AST mutants are then transformed back into source files for execution and testing. However, ESC mutation analysis cannot be performed the same way as for JavaScript programs. First, smart contracts must be deployed on a blockchain before being executed. Second, different from JavaScript, Solidity is not a scripting language. Thus, ESC must be compiled before being deployed. This significantly influences the process of ESC mutation analysis. Fig.~\ref{Performing-mutation-analysis-on-smart-contracts} illustrates how our ESC mutation analysis engine works. It treats the Smart Contract Under Test (SCUT) and a passed Test Suite (TS), which contains $n$ tests, as inputs, and finally outputs the mutation score. Below are the steps of mutation analysis on ESC. Note that steps (2) and (4) are different from the traditional JavaScript mutation testing process. 

\begin{itemize}
\item[(1)] \textbf{AST Generation}. SCUT is parsed into AST format for alleviating the loss of mutation precision, which may cause by the redundant statements, such as annotations and empty lines. To ensure the reliability of parsing result, solidity-parser-antlr \cite{federico2019solidity-parser-antlr}, a Solidity parser built on top of a robust ANTLR4 grammar, is selected in this step. 

\item[(2)] \textbf{AST Mutation}. The mutation is performed on the AST formatted SCUT, creating a new copy of the file for each AST mutant. In addition to thirteen new ESC mutation operators, which defined in \ref{Specific Mutation Operators}, we also reuse nine JavaScript-oriented operators \cite{mirshokraie2015guided} for ESC mutation. 

\item[(3)] \textbf{Mutant Generation and Filtration}. Each AST mutants will be transformed into a source file version, which has the identical semantic with SCUT excepting has been injected with a pre-defined fault. Subsequently, an equivalence checking is conducted to filter out the equivalent mutants. 

\item[(4)] \textbf{Compilation and Deployment}. ESC works on the Ethereum blockchain, and its execution result depends on the blockchain state, such as account balance and block height. Thus, to avoid the influence of blockchain state, for either SCUT or each source file version mutant, we should build a unique Ethereum testnet before its being compiled and deployed. Mutants that cause compilation errors are discarded immediately and not used in the following testing. 

\item[(5)] \textbf{SCUT and Mutants Testing}. Once an ESC has been successfully deployed, tests in TS can be executed. For each mutant, we record its execution results (i.e., passed or failed) in every test. Given a mutant $m_i$ and a test $t_j$, $m_i$ is marked as have been killed by $t_j$ if it failed on $t_j$.

\item[(6)] \textbf{Mutation Score Calculation}. After collecting all the execution results, we can get a $q \times n$ matrix $\ematrix$ of execution results, where $\ematrix_{i,j}$ depicts the execution result of mutant $m_i$ on test $t_j$. Finally, the mutation score is computed as a percentage of the mutants killed by the tests to the number of non-equivalent mutants. 

\end{itemize}

\section{Smart Contract Mutation Operators}
\label{Specific Mutation Operators}

To improve the effectiveness of mutation testing for smart contracts, we defined a new set of mutation operators for Ethereum smart contract. We read the Solidity documentation and look over the issues related to smart contracts on GitHub and Stack Exchange to design the mutation operators for the specific defects that may be generated by smart contracts. Details are as follows:

\subsection{Keyword Operators}
There are several keywords in Solidity that cannot be found in JavaScript. Six mutation operators are designed for these keywords.

\subsubsection{\textbf{Function State Keyword Change (FSC)}}
Functions can be declarable as \texttt{view}, which means that function will not modify the state of the contract, i.e. not modify variables, not emit events, etc. Functions can also be declarable as \texttt{pure}, which means that function can neither read from nor modify the state. Pure functions can only call other pure functions \cite{Github-pure-view}.

FSC operator change the state of a function by replace keyword \texttt{view} to \texttt{pure}. Table \ref{FSC} shows an example FSC mutant, if such mutant survives, it means you should use pure instead of view \textbf{(ESE\footnote{Ethereum Stack Exchange (ESE): https://ethereum.stackexchange.com/}\#28504)}.


\begin{table}[!htbp]%
	\centering 
	\caption{Example of FSC Mutant}
	\label{FSC}
	\begin{tabular}{|p{0.2cm}|p{7cm}|}
		\hline
		\emph{s$_{1}$}	&	\textbf{function func(uint x, uint y) \textcolor{red}{view} returns (uint)\{ }\\
        \emph{s$_{2}$}	&	\quad return x * (y + 42);\\
        \emph{s$_{3}$}	& \}\\
        \hline
        \hline
		\emph{s$_{1}$}	&	\textbf{function func(uint x, uint y) \textcolor{red}{pure} returns (uint)\{ }\\
        \emph{s$_{2}$}	&	\quad return x * (y + 42);\\
        \emph{s$_{3}$}	&	\}\\
        \hline
	\end{tabular}
\end{table}

\subsubsection{\textbf{Function Visibility Keyword Change (FVC)}}
Since Solidity has two kinds of function calls (\emph{internal} ones that do not create an actual EVM call and \emph{external} ones that do), there are four types of visibility for functions and state variables. \texttt{external} functions are part of the contract interface, which means they can be called from other contracts and via transactions. \texttt{public} functions are part of the contract interface and can be either called internally or via messages. \texttt{internal} functions can only be accessed from this contract and contracts deriving from it. \texttt{private} functions can only be visible for the contract where they are defined \textbf{(ESE\#32353)}. It is important to note that incorrect use of visibility keywords does not always lead to errors initially, but it can lead to faulty behavior when the contract is integrated with other classes, modified, or inherited from \cite{State-Variable-Default-Visibility}.

We design the FVC operator by imitating the Access Modifier Change operator in MuJava \cite{Ma2005}. Table \ref{FVC} shows an example FVC mutant where the visibility keyword \texttt{internal} is changed into \texttt{private}, which makes it impossible to call function \texttt{setData()} from  \texttt{C}'s derived contracts.


\begin{table}[!htbp]%
	\centering 
	\caption{Example of FVC Mutant}
	\label{FVC}
	\begin{tabular}{|p{0.2cm}|p{7cm}|}
		\hline
		\emph{s$_{1}$}	& contract C\{ \\
        \emph{s$_{2}$}	& \quad function f(uint a) private pure returns (uint b)\{\}\\
        \emph{s$_{3}$}	& \quad \textbf{function setData(uint a) \textcolor{red}{internal}\{data = a;\}}\\
        \emph{s$_{4}$}	& \quad uint public data;\\
        \emph{s$_{5}$}	& \}\\
        \hline
		\hline
		\emph{s$_{1}$}	& contract C\{ \\
        \emph{s$_{2}$}	& \quad function f(uint a) private pure returns (uint b)\{\}\\
        \emph{s$_{3}$}	& \quad \textbf{function setData(uint a) \textcolor{red}{private}\{data = a;\}}\\
        \emph{s$_{4}$}	& \quad uint public data;\\
        \emph{s$_{5}$}	& \}\\
        \hline
	\end{tabular}
\end{table}

\subsubsection{\textbf{Data Location Keyword Replacement (DLR)}}

Every reference type, i.e. arrays and structs, has an additional annotation in Solidity, the “data location”, about where it is stored. There are three data locations: \texttt{memory} (whose lifetime is limited to a function call), \texttt{storage} (the location where the state variables are stored and is persistent between function calls) and \texttt{calldata} (special data location that contains the function arguments, only available for external function call parameters) \cite{Solidity-Doc-Data-location}.

Data locations are not only relevant for the persistence of data, but also for the semantics of assignments. For example, Assignments between storage and memory (or from calldata) always create an independent copy while Assignments from memory to memory only create references \textbf{(ESE\#1231)}. Uninitialized local storage variables can point to unexpected storage locations in the contract, which can lead to intentional or unintentional vulnerabilities \cite{Uninitialized-Storage-Pointer}. Table \ref{DLR} shows an example DLR mutant where the data location of \texttt{mySandwich} is replaced from \texttt{storage} to \texttt{memory}. This mutant turns \texttt{mySandwich} from a reference into a copy. DLR mutants challenge testers to design test cases to check whether the original variables have changed.


\begin{table}[!htbp]%
	\centering 
	\caption{Example of DLR Mutant}
	\label{DLR}
	\begin{tabular}{|p{0.2cm}|p{7cm}|}
		\hline
		\emph{s$_{1}$}	& function eatSandwich(uint \_index) public\{ \\
        \emph{s$_{2}$}	& \quad \textbf{Sandwich \textcolor{red}{storage} mySandwich = sandwiches[\_index];}\\
        \emph{s$_{3}$}	& \quad mySandwich.status = ``Eaten!'';\\
        \emph{s$_{4}$}	& \}\\
        \hline
		\hline
		\emph{s$_{1}$}	& function eatSandwich(uint \_index) public\{ \\
        \emph{s$_{2}$}	& \quad \textbf{Sandwich \textcolor{red}{memory} mySandwich = sandwiches[\_index];}\\
        \emph{s$_{3}$}	& \quad mySandwich.status = ``Eaten!'';\\
        \emph{s$_{4}$}	& \}\\
        \hline
	\end{tabular}
\end{table}

\subsubsection{\textbf{Variable Type Keyword Replacement (VTR)}}
Since Solidity is a statically and strongly typed language, the type of each variable needs to be specified. Solidity supports three fixed-size types, fixed-size integers, fixed-point numbers and fixed-size byte arrays. All these types are restricted to a certain range. For example, an 8-bit unsigned integer can store values between 0 and 255 (2$^8$-1). When the result of some arithmetic falls outside that supported range, an overflow occurs \textbf{(ESE\#7293)}. 

Integers in Solidity are divided into signed integers and unsigned integers of various sizes. The consequence of an integer overflow is that the most significant bits of the result are lost, which can cause real-world vulnerabilities such as \textbf{batchOverflow}\cite{batchOverflow}. Fixed point numbers and fixed-size byte arrays are no difference. To ensure that the correct value keywords are used, we define a series of VTR operators such as replace \texttt{uint} to \texttt{int} and replace \texttt{bytes32} to \texttt{bytes8}. Table \ref{VTR} shows an example of VTR mutant. This kind of mutants requires testers to consider negative number and truncation.


\begin{table}[!htbp]%
	\centering 
	\caption{Example of VTR Mutant}
	\label{VTR}
	\begin{tabular}{|p{0.2cm}|p{7cm}|}
		\hline
		\emph{s$_{1}$}	& \textbf{for (\textcolor{red}{uint256} i=0; i<length; i++)\{} \\
        \emph{s$_{2}$}	& \quad A.push(B[i]);\\
        \emph{s$_{3}$}	& \}\\
        \hline
		\hline
		\emph{s$_{1}$}	& \textbf{for (\textcolor{red}{uint8} i=0; i<length; i++)\{} \\
        \emph{s$_{2}$}	& \quad A.push(B[i]);\\
        \emph{s$_{3}$}	& \}\\
        \hline
	\end{tabular}
\end{table}

\subsubsection{\textbf{Payable Keyword Deletion (PKD)}}
\texttt{payable} is a modifier that allows a function to be called with a non-zero value (that you can access via msg.value). If a function needs currency operation, it must have a \texttt{payable} keyword, so that it can receive Ethernet currency normally \textbf{(ESE\#20847)}. The PKD operator deletes the \texttt{payable} keyword of a function at a time, and then any transaction trying to send ether will be rejected. The sample program is shown in Table \ref{PKD}.


\begin{table}[!htbp]%
	\centering 
	\caption{Example of PKD Mutant}
	\label{PKD}
	\begin{tabular}{|p{0.2cm}|p{7cm}|}
		\hline
		\emph{s$_{1}$}	& \textbf{function deposit() \textcolor{red}{payable}\{} \\
        \emph{s$_{2}$}	& \quad deposits[msg.sender] += msg.value;\\
        \emph{s$_{3}$}	& \}\\
        \hline
		\hline
		\emph{s$_{1}$}	& \textbf{function deposit()\{} \\
        \emph{s$_{2}$}	& \quad deposits[msg.sender] += msg.value;\\
        \emph{s$_{3}$}	& \}\\
        \hline
	\end{tabular}
\end{table}

\subsubsection{\textbf{Delete Keyword Deletion (DKD)}}
In Solidity, \texttt{delete a} assigns the initial value for the type to \texttt{a}. For integers, it is equivalent to \texttt{a = 0}, but it can also be used on arrays, where it assigns a dynamic array of length zero or a static array of the same length with all elements set to their initial values \textbf{(ESE\#58495)}. It is important to note that deleting \texttt{a}  really behaves like an assignment to \texttt{a}, i.e. it stores a new object in \texttt{a}. Table \ref{DKD} shows an example of DKD mutant which makes the return value change from 0 to 3. The DKD operator is similar to the Member Variable Initialization Deletion (JID) in MuJava \cite{Ma2005}.


\begin{table}[!htbp]%
	\centering 
	\caption{Example of DKD Mutant}
	\label{DKD}
	\begin{tabular}{|p{0.2cm}|p{7cm}|}
		\hline
		\emph{s$_{1}$}	& function testDel() returns (uint)\{ \\
        \emph{s$_{2}$}	& \quad uint a = 3;\\
        \emph{s$_{3}$}	& \quad \textbf{\textcolor{red}{delete a;}}\\
        \emph{s$_{4}$}	& \quad return a;\\
        \emph{s$_{5}$}	& \}\\
        \hline
		\hline
		\emph{s$_{1}$}	& function testDel() returns (uint)\{ \\
        \emph{s$_{2}$}	& \quad uint a = 3;\\
        \emph{s$_{3}$}	& \quad \textbf{\textcolor{red}{//delete a;}}\\
        \emph{s$_{4}$}	& \quad return a;\\
        \emph{s$_{5}$}	& \}\\
        \hline
	\end{tabular}
\end{table}

\subsection{Global Variables and Functions Operators}
There are some special variables and functions which always exist in the global namespace \cite{Solidity-Special-Variables-and-Functions}. Thus, we define three mutation operators for them. 

\subsubsection{\textbf{Global Variable Change (GVC)}}
Solidity uses several global variables to provide information about the blockchain \textbf{(ESE\#2664)}. For example, \texttt{now} is the current block timestamp (alias for \texttt{block.timestamp}), \texttt{block.number} shows the current block number, and \texttt{msg.value} shows the number of wei sent to the contract. A GVC operator changes a global variable by assigning it a format-compliant random value, and thus may cause different execution results. The sample program is shown in Table \ref{GVC}, testers can kill the mutant by executing function \texttt{getNow()}.


\begin{table}[!htbp]%
	\centering 
	\caption{Example of GVC Mutant}
	\label{GVC}
	\begin{tabular}{|p{0.2cm}|p{7cm}|}
		\hline
		\emph{s$_{1}$}	& function getNow() public constant returns (uint)\{ \\
        \emph{s$_{2}$}	& \quad \textbf{return \textcolor{red}{now};}\\
        \emph{s$_{3}$}	& \}\\
        \hline
		\hline
		\emph{s$_{1}$}	& function getNow() public constant returns (uint)\{ \\
        \emph{s$_{2}$}	& \quad \textbf{return \textcolor{red}{0};}\\
        \emph{s$_{3}$}	& \}\\
        \hline
	\end{tabular}
\end{table}

\subsubsection{\textbf{Mathematical Functions Replacement (MFR)}}
\texttt{addmod} and \texttt{mulmod} are two mathematical global functions in Solidity. \texttt{addmod(uint x, uint y, uint k)} computes $(x + y) \% k$ where the addition is performed with arbitrary precision and does not wrap around at 2**256. Similarly, \texttt{mulmod(uint x, uint y, uint k)} computes $(x * y) \% k$ \cite{Solidity-Mathematical-Functions}. We design MFR operators to exchange them to keep developers from accidentally using incorrect functions. Table \ref{MFR} shows an example of MFR mutant, testers can kill the mutant by testing the return value of the function.


\begin{table}[!htbp]%
	\centering 
	\caption{Example of MFR Mutant}
	\label{MFR}
	\begin{tabular}{|p{0.2cm}|p{7cm}|}
		\hline
		\emph{s$_{1}$}	& function Test(uint x, uint y, uint k) view public returns (uint)\{ \\
        \emph{s$_{2}$}	& \quad \textbf{return \textcolor{red}{addmod}(x, y, k);}\\
        \emph{s$_{3}$}	& \}\\
        \hline
		\hline
		\emph{s$_{1}$}	& function Test(uint x, uint y, uint k) view public returns (uint)\{ \\
        \emph{s$_{2}$}	& \quad \textbf{return \textcolor{red}{mulmod}(x, y, k);}\\
        \emph{s$_{3}$}	& \}\\
        \hline
	\end{tabular}
\end{table}

\subsubsection{\textbf{Address Variable Replacement (AVR)}}
There are three global variables related to address in Solidity. \texttt{block.coinbase} is the current block miner’s address, \texttt{msg.sender} is the sender of the message (current call), and \texttt{tx.origin} is the sender of the transaction (full call chain). The different between \texttt{msg.sender} and \texttt{tx.origin} is that \texttt{msg.sender} can be a contract but \texttt{tx.origin} can never be a contract. In a simple call chain, $A \to B \to C \to D$, \texttt{msg.sender} in D will be C, and \texttt{tx.origin} will always be A \textbf{(ESE\#1891)}. 

An AVR operator replaces an address variable with another one. For example, in Table \ref{AVR}, \texttt{msg.sender} is replaced with \texttt{tx.origin}. To kill this mutant, a test case needs to be designed where \texttt{msg.sender} is not equal to \texttt{tx.origin}.


\begin{table}[!htbp]%
	\centering 
	\caption{Example of AVR Mutant}
	\label{AVR}
	\begin{tabular}{|p{0.2cm}|p{7cm}|}
		\hline
		\emph{s$_{1}$}	& function sendTo(address receiver, uint amount) public\{ \\
        \emph{s$_{2}$}	& \quad \textbf{require(\textcolor{red}{msg.sender} == owner);}\\
        \emph{s$_{3}$}	& \quad receiver.transfer(amount);\\
        \emph{s$_{4}$}	& \}\\
        \hline
		\hline
		\emph{s$_{1}$}	& function sendTo(address receiver, uint amount) public\{ \\
        \emph{s$_{2}$}	& \quad \textbf{require(\textcolor{red}{tx.origin} == owner);}\\
        \emph{s$_{3}$}	& \quad receiver.transfer(amount);\\
        \emph{s$_{4}$}	& \}\\
        \hline
	\end{tabular}
\end{table}

\subsection{Variable Unit Operators}
Solidity has two unique variable units, Ether Units and Time Units. We defined the following two mutation operators to simulate the defects caused by using the wrong unit.

\subsubsection{\textbf{Ether Unit Replacement (EUR)}}
Units are indispensable whether it is to construct transactions for the transfer of Ethernet currency or to invoke intelligent contracts for the issuance of tokens. Ether's unit suffixes are \texttt{wei}, \texttt{finney}, \texttt{szabo}, \texttt{ether}. The conversion format is $1 ether = 1 * 10^3 finney = 1 * 10^6 szabo = 1 * 10^18 wei$ \cite{Solidity-Ether-Units}. Table \ref{EUR} shows an EUR mutant where the unit keyword \texttt{finney} has been replaced by \texttt{szabo}. To kill this mutant, \texttt{this.balance} should be set between 70 \texttt{finney} and 70 \texttt{szabo}.


\begin{table}[!htbp]%
	\centering 
	\caption{Example of EUR Mutant}
	\label{EUR}
	\begin{tabular}{|p{0.2cm}|p{7cm}|}
		\hline
		\emph{s$_{1}$}	& \textbf{if(this.balance \textgreater= 70 \textcolor{red}{finney})\{} \\
        \emph{s$_{2}$}	& \quad uint sendProfit = this.balance;\\
        \emph{s$_{3}$}	& \}\\
        \hline
		\hline
		\emph{s$_{1}$}	& \textbf{if(this.balance \textgreater= 70 \textcolor{red}{szabo})\{} \\
        \emph{s$_{2}$}	& \quad uint sendProfit = this.balance;\\
        \emph{s$_{3}$}	& \}\\
        \hline
	\end{tabular}
\end{table}

\subsubsection{\textbf{Time Unit Replacement (TUR)}}
Time is also one of the characteristics of smart contracts. The time units supported in solidity are \texttt{seconds}, \texttt{minutes}, \texttt{hours}, \texttt{days} and \texttt{weeks}, where \texttt{seconds} is the default unit \cite{Solidity-Time-Units}. The TUR operator replace a time unit suffix with another one, the sample program is shown in Table \ref{TUR}.


\begin{table}[!htbp]%
	\centering 
	\caption{Example of TUR Mutant}
	\label{TUR}
	\begin{tabular}{|p{0.2cm}|p{7cm}|}
		\hline
		\emph{s$_{1}$}	& function f(uint start, uint timeAfter)\{ \\
        \emph{s$_{2}$}	& \quad \textbf{if (block.timestamp \textgreater= start + timeAfter * 1 \textcolor{red}{days})\{\}}\\
        \emph{s$_{3}$}	& \}\\
        \hline
		\hline
		\emph{s$_{1}$}	& function f(uint start, uint timeAfter)\{ \\
        \emph{s$_{2}$}	& \quad \textbf{if (block.timestamp \textgreater= start + timeAfter * 1 \textcolor{red}{weeks})\{\}}\\
        \emph{s$_{3}$}	& \}\\
        \hline
	\end{tabular}
\end{table}

\subsection{Error Handling Operators}
Solidity uses state-reverting exceptions to handle errors. Such an exception will undo all changes made to the state in the current call. The convenience functions assert and require can be used to check for conditions and throw an exception if the condition is not met \cite{Solidity-Error-handling}. We define two kinds of mutation operators related to error handling.

\subsubsection{\textbf{Require Statement Deletion (RSD)}}
The \texttt{require} function is often used to ensure valid conditions on inputs or contract state variables, or to validate return values from calls to external contracts. If the condition in \texttt{require()} is not satisfied, the state change is revoked to check for errors caused by input or external components, and an error message can be provided at the same time. The RSD operator is inspired from the mutation operators related to \texttt{if} statement, but the difference is that \texttt{require} reverts the entire state changes in the function while \texttt{if} doesn’t \textbf{(ESE\#60585)}.

As is shown in Table \ref{RSD}, the RSD operator deletes the whole \texttt{require} statement by commenting them out to ensure that the following statements always execute. In addition, change the statement in \texttt{require()} to \texttt{false} is another operator, named \emph{\textbf{Require Statement Change (RSC)}}, which ensures the following statement will never get executed.

\subsubsection{\textbf{Assert Statement Deletion (ASD)}}
The \texttt{assert} function works in a similar way to \texttt{require}, but \texttt{require} is used to check conditions on inputs while \texttt{assert} is used for internal error checking \textbf{(ESE\#16457)}. The \texttt{assert} statement operators are also divided into two kinds. The ASD operator can enforce the execution of the following statements by deleting the \texttt{assert} statement (for example, line \emph{s$_5$} in Fig. \ref{RSD}). Relatively, the \emph{\textbf{Assert Statement Change (ASC)}} operator can make the following statements never execute by changing the condition statement to \texttt{false}.


\begin{table}[!htbp]%
	\centering 
	\caption{Example of RSD Mutant}
	\label{RSD}
	\begin{tabular}{|p{0.2cm}|p{7cm}|}
		\hline
		\emph{s$_{1}$}	& function sendHalf(address a) public payable returns (uint)\{ \\
        \emph{s$_{2}$}	& \quad \textbf{\textcolor{red}{require(msg.value\%2 == 0, ``Even value required.'');}}\\
        \emph{s$_{3}$}	& \quad uint balanceBefore == this.balance;\\
        \emph{s$_{4}$}	& \quad addr.transfer(msg.value / 2);\\
        \emph{s$_{5}$}	& \quad assert(this.balance == balanceBefore - msg.value / 2);\\
        \emph{s$_{6}$}	& \quad return this.balance;\\
        \emph{s$_{4}$}	& \}\\
        \hline
		\hline
		\emph{s$_{1}$}	& function sendHalf(address a) public payable returns (uint)\{ \\
        \emph{s$_{2}$}	& \quad \textbf{\textcolor{red}{//require(msg.value\%2 == 0, ``Even value required.'');}}\\
        \emph{s$_{3}$}	& \quad uint balanceBefore == this.balance;\\
        \emph{s$_{4}$}	& \quad addr.transfer(msg.value / 2);\\
        \emph{s$_{5}$}	& \quad assert(this.balance == balanceBefore - msg.value / 2);\\
        \emph{s$_{6}$}	& \quad return this.balance;\\
        \emph{s$_{4}$}	& \}\\
        \hline
	\end{tabular}
\end{table}

\subsection{Summary}


In summary, as depicted in table \ref{Specific-mutation-operators}, we proposed 15 ESC specific mutation operators, including six corresponds to keyword, three corresponds to global variable/function, two corresponds to variable unit, and four corresponds to error handling in Ethereum smart contract. 

\begin{table}[htbp]
	\centering 
	\caption{Specific Mutation Operators}
	\label{Specific-mutation-operators}
	\begin{tabular}{c|c|l}
		\toprule
		\textbf{Type} & \textbf{Operator} & \textbf{Description}\\
		\hline
		\multirow{6}{*}{Keyword}    
		                            &FSC    & Function State Keyword Chang \\
		                            &FVC    & Function Visibility Keyword Chang \\ 
		                            &DLR    & Data Location Keyword Replacement \\ 
		                            &VTR    & Variable Type Keyword Replacement \\ 
		                            &PKD    & Payable Keyword Deletion \\ 
		                            &DKD    & Delete Keyword Deletion \\ \hline
		\multirow{3}{*}{\tabincell{c}{ Global Variable\\and Function}}  
		                            &GVC    & Global Variable Change \\ 
		                            &MFR    & Mathematical  Functions  Replacement \\ 
		                            &AVR    & Address  Variable  Replacement \\ \hline
		\multirow{2}{*}{Variable Unit}  
		                            &EUR    & Ether Unit Replacement \\ 
		                            &TUR    & Time Unit Replacement \\ \hline
	    \multirow{4}{*}{Error Handling}  
	                                &RSD    & Require Statement Deletion \\ 
		                            &RSC    & Require Statement Change \\ 
		                            &ASD    & Assert Statement Deletion \\ 
		                            &ASC    & Assert Statement Change \\ 
		\bottomrule
	\end{tabular}
\end{table}

\section{Empirical Study}

In this section, we assess the proposed mutation testing approach and the value of each mutation operator for ESC. To that end, we first present the addressed research questions followed by the experimental subjects and experiments performed. 

\subsection{Research Questions}

The goal of this section is to answer the following research questions:

\begin{itemize}
	\item \textbf{RQ1: Are mutation testing effective in evaluating the adequacy of ESC test-suite?} For a test-suite, its test adequacy can be measured by its defect detect capability. We aim to verify that mutation testing performs well in measuring the adequacy of ESC test-suite, and expect mutation testing is stronger than coverage based approach. 
	\item \textbf{RQ2: Are the specific mutation operators work well in mutation testing for ESC?} For each specific mutation operators, we intend to know (1) the non-equivalent mutant generation rate when applying it in ESC mutation testing and (2) if they lead to real bugs in practice. This fact would allow us to analyze the correlation between mutants and real defects.  
\end{itemize}

\subsection{Experimental Subjects}

\begin{table}[!htbp]%
	\centering 
	\caption{Experimental Subjects}
	\label{Experimental-Subject}
	\begin{tabular}{cccc}
		\toprule
		\textbf{DApp} 	& \textbf{LOC} & \textbf{BOC} & \textbf{STS} \\
		\midrule
		SkinCoin	  &225	&66	  	  & 35	\\
		SmartIdentity &180  &34       & 87	\\
		AirSwap       &330  &76       & 17 	\\
		CryptoFin     &348  &58       & 44	\\
		\bottomrule
	\end{tabular}
\end{table}

Our study regards a set of smart contracts in four different real-world Ethereum DApps (i.e., SkinCoin~\cite{SkinCoin2019}, SmartIdentity~\cite{SmartIdentity2019}, AirSwap~\cite{AirSwap2019} and CryptoFin~\cite{CryptoFin2019}) as experimental subjects. SkinCoin is a universal cryptocurrency for instant trading skins in games and making bets on e-sports events. SmartIdentity relies on Ethereum blockchain to represent an identity using a smart contract. CryptoFin is a collection of Solidity libraries, with an initial focus on arrays. AirSwap is a peer-to-peer trading network built on Ethereum. These DApps are selected because each of which provides not only a set of ESCs but also accompanies a well-designed test-suite. Thus, we do not need to manually design an ESC test-suite, which is too subjective to generate a fair experimental result and conclusion. Tools for automated ESC test generating such as ContractFuzzer~\cite{jiang2018} and MTG~\cite{Wang2019} are also not used because the generated results are not complete, i.e., missing assert statements. Tests that failed on the original DApp are removed from the subject (e.g., we removed 15 tests from AirSwap's test-suite, which contains 32 tests). Table \ref{Experimental-Subject} summarizes the characteristics of DApps used in the experiments. For each DApp, its name (col. 1), lines of code (col. 2), branches of code (col. 3), and size of test-suite (col. 4) are described. All the smart contracts have been open sourced on GitHub. 

\subsection{Experiment 1: Mutation Testing Effectiveness}

In this experiment, we compare mutation testing to coverage based approach for evaluating the effectiveness of mutation testing. 

\subsubsection{Setup}

\textbf{Mutants Generation.} Given DApp under test, we first generated all the possible mutants by using the total of 25 mutation operators, of which 10 are general mutation operators, and 15 are proposed by us. After removing the equivalent mutants and those were failed during compiling, we got $n$ useful mutants. \textbf{Mutants Partition and Execution.} Then, we randomly selected $\frac{n}{2}$ useful mutants to build mutant set $M_1$, and the remaining useful mutants build mutant set $M_2$. Each mutant in $M_1$ was tested by the accompanying Test-Suite $TS=\{ t_i|i \in [ 1,n ] \}$. $\textbf{TS}_{Cov}$~\textbf{Generation.}~$TS_{Cov}$ was generated by randomly selecting a subset of $TS$ that satisfies test targets w.r.t. line coverage and branch coverage, i.e., line coverage and branch coverage of $TS_{Cov}$ are equal to that of $TS$ respectively. $\textbf{TS}_{M_1}$~\textbf{Generation.}~${TS}_{M_1}$ was generated by randomly selecting a subset of $TS$ that had the same mutation score with $TS$ on $M_1$. \textbf{Comparison} Regarding $M_2$ as the verification set and finally calculating the mutation scores of $TS_{Cov}$ and ${TS}_{M_1}$ on $M_2$ for comparison. For accurately, we ran the analysis ten times to obtain the results on average to reduce bias introduced by randomization.

\subsubsection{Results}

\begin{table}[!htbp]
	\centering 
	\caption{Mutant Number of Each DApp}
	\label{Mutant-Number-of-Each-DApp}
	\begin{tabular}{|c|c|c|c|c|c|c|}
		\hline
		\multirow{2}*{\textbf{DApp}} & \multicolumn{2}{c|}{\textbf{General}} & \multicolumn{2}{c|}{\textbf{ESC Specific}} & 
		\multicolumn{2}{c|}{\textbf{Total}} \\
		\cline{2-7}
		                 & \textbf{ALL} & \textbf{NEQ} & \textbf{ALL} & \textbf{NEQ} & \textbf{ALL} & \textbf{NEQ}\\
		\hline
		SkinCoin         &314  &274  &106  &106  &420  &380 \\
		\hline
		SmartIdentity    &267  &225  &86   &86   &353  &311 \\
		\hline
		AirSwap          &406  &363  &175  &175  &581  &538 \\
		\hline
		Cryptofin        &464  &406  &214  &183  &678  &589 \\
		\hline
		\textbf{Total}   &1451 &1268 &581  &550  &2032 &1818 \\
		\hline
	\end{tabular}
\end{table}

Table \ref{Mutant-Number-of-Each-DApp} summarizes the generated mutants. For each DApp, numbers of mutants (col. 2, 4 and 6) that were generated by general operators, ESC specific operators and total operators are presented, together with the ones of non-equivalent mutants (col. 3, 5 and 7). The former five columns show an 84.27\% to 89.41\% non-equivalent mutating rate for general operators and a comparable 85.50\% to 100\% non-equivalent mutating rate for ESC specific operators, indicating the proposed operators will not increase the cost of detecting equivalent mutants during mutation testing. In total, 1818 mutants were selected from the mutant pool, which contains 2032 compilable mutants, to conduct the first experiment. 

\begin{table*}[t]%
	\centering 
	\caption{Coverage Results and Mutation Scores of $TS$, $TS_{Cov}$ and $TS_{MS_1}$}
	\label{Coverage-Results-and-Mutation-Scores}
	\begin{tabular}{|c|c|c|c|c|c|c|c|c|c|c|c|c|}
		\hline
		\multirow{2}*{\textbf{DApp}} & \multicolumn{4}{c|}{\textbf{TS}} & \multicolumn{4}{c|}{\textbf{$TS_{Cov}$}} & \multicolumn{4}{c|}{\textbf{$TS_{MS_1}$}} \\
		\cline{2-13} 
        & \textbf{LCov} & \textbf{BCov} & $\textbf{MS}_{1}$ & $\textbf{MS}_{2}$
        & \textbf{LCov} & \textbf{BCov} & $\textbf{MS}_{1}$ & $\textbf{MS}_{2}$ 
        & \textbf{LCov} & \textbf{BCov} & $\textbf{MS}_{1}$ & $\textbf{MS}_{2}$ \\
        \hline
        SkinCoin
        & 80.44 & 63.33 & 34.7  & 43.1  
        & -     & -     & 15.2  & 23.6
        & 79.33 & 59.98 & -     & 40.3 \\
        \hline
        SmartIdentity
        & 91.67 & 85.29 & 63.6  & 47.4 
        & -     & -     & 25.7  & 23.5
        & 91.67 & 85.29 & -     & 44.6 \\
        \hline
        AirSwap
        & 45.45 & 17.11 & 32.2  & 28.8 
        & -     & -     & 19.3  & 17.6
        & 45.45 & 17.11 & -     & 28.8 \\
        \hline
        Cryptofin
        & 89.78 & 76.47 & 47.8  & 53.2 
        & -     & -     & 28.4  & 31.3
        & 88.71 & 76.47 & -     & 51.7 \\
        \hline
        \textbf{Average}
        & 76.84 & 60.55 & 44.6 & 43.1 
        & -     & -     & 22.2 & 24.0
        & 76.29 & 59.71 & -    & 41.4 \\
		\hline
	\end{tabular}
\end{table*}

Table \ref{Coverage-Results-and-Mutation-Scores} presents the results of line coverage and branch coverage of $TS$ (col. 2-3) and $TS_{Cov}$ (col. 10-11) when applying ESC under test, as well as the mutation scores of $TS$ (col. 4-5), $TS_{Cov}$ (col. 8-9) and $TS_{MS_1}$ (col. 13) when applying ${MS}_{1}$ and ${MS}_{2}$. For the mutants that were killed by $TS$, $TS_{MS_1}$ can kill most of them (i.e., $96.01\%=\frac{{MS}_{2}(TS_{MS_1})}{{MS}_{2}(TS)}$ in average), whereas $TS_{Cov}$ can only kill half of them (i.e., $55.68\%=\frac{{MS}_{2}(TS_{Cov})}{{MS}_{2}(TS)}$ in average), indicating that $TS_{MS_1}$ outperforms $TS_{Cov}$ in defect detection. To determine whether the observed difference is statistically significant or not, we applied the paired Wilcoxon test and carried out the two-tailed alternative hypothesis. The value of the test is 0.005. Therefore, we can accept the alternative hypothesis that $TS_{MS_1}$ significantly outperforms $TS_{Cov}$ and can make a conclusion that mutation testing is more effective in evaluating the adequacy of test-suite. 

\subsection{Experiment 2: Mutation Operator Effectiveness}

In this experiment, we assessed each mutation operator by rates of survival and equivalent. We also conducted a survey on open-sourced communities to analysis the corresponds between real defects and mutants.

\subsubsection{Setup}

\textbf{Mutants Classification.} For both general mutants and ESC specific mutants, we further classified them and their testing results based on their mutation operators. Besides experiment, we inspected real defect reports for evaluating operator effectiveness. \textbf{Defect Reports Collection.} We searched the defect reports from GitHub~\cite{Examples-of-Solidity-Security-Issues, Smart-Contract-Weakness-Classification-and-Test-Cases, USCC-Submissions-2017, Ethereum-Solidity-Issues, Ethereum-GoEthereum-Issues}, DASP~\cite{Decentralized-Application-Security-Project} and PeckShield~\cite{PeckShield}, and finally collected 729 closed reports w.r.t. ESC. For each report, we determined it into one type of mutation operator. 

\begin{table}[t]%
    \setlength{\abovedisplayskip}{-0.2cm}
    \setlength{\belowdisplayskip}{-0.5cm}
	\centering 
	\caption{Experimental Results of Mutation Operators}
	\label{Statistics-for-Each-Mutation-Operator}
	\begin{tabular}{|p{1.2cm}<{\centering}|p{0.8cm}<{\centering}|p{0.8cm}<{\centering}|p{0.8cm}<{\centering}|p{0.8cm}<{\centering}|p{0.8cm}<{\centering}|}
		\hline
    	\multirow{2}*{\textbf{Operator}} & \multicolumn{4}{c|}{\textbf{Number of Mutants}} & \multirow{2}*{\textbf{MS}} \\
		\cline{2-5}
		                                 & \textbf{All} & \textbf{Equ.} & \textbf{Killed} & \textbf{Live} &  \\
		\hline
		\hline
		\multicolumn{6}{|c|}{\textbf{General Mutants}} \\ \hline
		AOR      & 257    & 4      &117     & 136    & 46.2 \\ \hline
		AOI      & 416    & 108    & 128    & 159    & 44.6 \\ \hline
		ROR      & 410    & 41     & 148    & 221    & 40.1 \\ \hline
		COR      & 29     & 0      & 5      & 24     & 17.2 \\ \hline
		LOR      & 0      & 0      & 0      & 0      &  0.0 \\ \hline
		ASR      & 34     & 0      & 18     & 16     & 52.9 \\ \hline
		SDL      & 229    & 30     & 78     & 121    & 39.2 \\ \hline
		RVR      & 36     & 0      & 11     & 25     & 30.6 \\ \hline
		CSC      & 40     & 0      & 26     & 14     & 65.0 \\ \hline
		Subtotal & 1451   & 183    & 531    & 716    & 42.6 \\ 
		\hline
		\hline
		\multicolumn{6}{|c|}{\textbf{ESC Specific Mutants}} \\ \hline
		FSC      & 4      & 0      & 0      & 4      &  0.0 \\ \hline
		FVC      & 136    & 0      & 67     & 69     & 49.3 \\ \hline
		DLR      & 6      & 0      & 1      & 5      & 16.7 \\ \hline
		VTR      & 101    & 31     & 16     & 54     & 22.9 \\ \hline
		PKD      & 1      & 0      & 0      & 1      &  0.0 \\ \hline
		DKD      & 6      & 0      & 2      & 4      & 33.3 \\ \hline
		GVC      & 65     & 0      & 34     & 31     & 52.3 \\ \hline
		MFR      & 55     & 0      & 19     & 36     & 34.5 \\ \hline
		AVR      & 127    & 0      & 35     & 103    & 25.4 \\ \hline
		EUR      & 20     & 0      & 4      & 16     & 20.0 \\ \hline
		TUR      & 6      & 0      & 0      & 6      &  0.0 \\ \hline
		RSD      & 13     & 0      & 3      & 10     & 23.1 \\ \hline
		RSC      & 13     & 0      & 5      & 8      & 38.5 \\ \hline
		ASD      & 19     & 0      & 2      & 17     & 10.5 \\ \hline
		ASC      & 19     & 0      & 13     & 6      & 68.4 \\ \hline
		Subtotal & 581    & 31     & 201    & 370    & 35.2 \\ 
		\hline
		\hline
		\textbf{Total}&\textbf{2032}&\textbf{214}&\textbf{732}&\textbf{1086}&\textbf{40.3}\\ \hline
	\end{tabular}
\end{table}

\subsubsection{Results} Table~\ref{Statistics-for-Each-Mutation-Operator} presents the statistics for each mutation operator, where columns 2-5 respectively depicts the numbers of all, equivalent, killed and live mutants and the last column depicts the mutation score. The first group is the mutants generated by general mutation operators. In addition to LOR, other general mutation operators generate at least one mutant. Among them, COR has the lowest mutation score of 17.2, but this only shows that the test cases are not sufficient in this respect, and does not mean that COR operator is ineffective. AOI generates the most mutants. However, 108 of the 416 mutants generated by AOI are equivalent, this is because the inserted arithmetic operators do not work on variables that have an impact on the execution results. The 10 traditional mutation operators generate a total of 1451 mutants, of which 1247 were non-equivalent, so the non-equivalent mutant generation rate is 85.94\%.

The second group corresponds to ESC specific mutation operators. FSC, PKD and TUR have the lowest mutation scores of 0, indicating that none of their mutants was killed by $TS$. These three operators only generated eleven mutants, so this low percentage probably isn't meaningful. FVC and AVR generated 136 and 127 mutants respectively, accounting for nearly half of all mutants. In addition, VTR generated 101 mutants, but 31 of them are equivalent mutants. For example, VTR can replace \texttt{uint} in \texttt{(uint i = 0; i < 20; i++)} with \texttt{int} and \texttt{uint8}, thus result in two mutants. However, both mutants will not affect the result of program execution, so they are logically equivalent. ASC generated 19 mutants and got the highest mutation score of 68.4, while ASD generates 19 mutants with the lowest mutation score of 10.5. This probably indicates that many test cases only trigger \texttt{assert(true)} without considering \texttt{assert(false)}, which may lead to a hidden defect. The average mutation score of the special mutation operators on $TS$ is 35.2, which is lower than that of the general mutation operators (42.6). This shows that testers pay less attention to the problems caused by solidity characteristics when writing test cases. This reflects the significance of our mutation operators.

Further, we conducted a survey to look up various issues and bug reports related to ESC from open source communities. Among 729 reports, 117 are related to our mutation operator, including 41 Keyword Operators bugs, 35 Global Variables and Functions Operators bugs, 9 Variable Unit Operators bugs and 32 Error Handling Operators bugs. Representative samples are as follows:

\begin{itemize}
	\item \textbf{SWC-100 (FVC)} Functions that do not have a function visibility type specified are public by default. This can lead to a vulnerability if a developer forgot to set the visibility and a malicious user is able to make unauthorized or unintended state changes.
	\item \textbf{SWC-109 (DLR)} Uninitialized local storage variables can point to unexpected storage locations in the contract, which can lead to unintentional vulnerabilities.
	\item \textbf{DASP\#item-3 (VTR/MFR)} An overflow condition gives incorrect results and, particularly if the possibility has not been anticipated, can compromise a program’s reliability and security.	\item \textbf{SWC-120 (GVC)} \texttt{block.timestamp} is insecure, as a miner can choose to provide any timestamp within a few seconds and still get his block accepted by others. Use of \texttt{blockhash}, \texttt{block.difficulty} and other fields is also insecure, as they're controlled by the miner.
	\item \textbf{DASP\#item-2 (AVR/RSD/RSC)} Access Control vulnerabilities can occur when contracts use \texttt{tx.origin} instead of \texttt{msg.sender} to validate callers, handle large authorization logic with lengthy \texttt{require} and make reckless use of \texttt{delegatecall} in proxy libraries or proxy contracts.
	\item \textbf{SWC-123 (ASD/ASC)} The Solidity \texttt{assert()} function is meant to assert invariants. Properly functioning code should never reach a failing assert statement. A reachable assertion mean that a bug exists in the contract that allows it to enter an invalid state or the assert statement is used incorrectly (e.g. to validate inputs).
\end{itemize}

The result shows that our newly proposed Solidity mutation operator can inject real defects, and thus can help developers avoid some common mistakes.

\section{Threats to Validity}

\textbf{Internal validity.} There are two main internal threats to validity. Firstly, whether we mutate the smart contract correctly. Our mutation operations are implemented at the AST level, where AST is generated with solidity-parser-antlr, the mutation and restoration of AST are implemented by ourselves. To verify that our operators were correctly implemented, we checked every mutant by hand to ensure it was mutated as expected. Secondly, due to the constant update of the solidity version, the parser and mutation tool need to be updated accordingly. Meanwhile, the design of mutation operators should also consider the impact of the version update to meet the latest test requirements.

\textbf{Construct validity.} There are two construct threats to validity. Firstly, the test code of all smart contracts is attached to original the project, while the quality of test cases will have an impact on our experiment (for example, low-quality test cases will make all the generated mutants survive). Secondly, the identification of equivalent mutants is performed manually, so we can not ensure that all equivalent mutants are excluded.

\textbf{External validity.} Our experiments were conducted on 26 smart contracts from four DApps, so it is not possible to guarantee the representative of selected subjects. We managed to select DApps from four different areas to maximize the representation of the experiment.

\section{Related Work}

In this section, we will describe related work in two areas: Mutation Testing and Smart Contract Testing.

\subsection{Mutation Testing}

Mutation testing was first discovered and made public by DeMillo, Lipton and Sayward \cite{Demillo1978}, and explored 
extensively by Offutt and others \cite{Offutt2001}.
It is based on two premises: 
the competent programmer hypothesis \cite{Demillo1978} and the coupling effect hypothesis \cite{Offutt1992}. The skilled programmer hypothesis assumes that 
the defect code written by the programmer is very close to the correct code, and that the defect can be removed with only 
minor modifications. Based on this assumption, mutation testing can simulate the actual programming behavior of skilled 
programmers only by modifying the amplitude code of the program under test. Coupling effect \cite{Offutt1992} hypothesis 
points out that complex faults are coupled with simple faults, so a test data set that detects simple faults (such as those 
introduced by mutation) will detect complex faults, i.e., the combination of several simple faults.

Mutation testing has been applied to many programming languages such as C \cite{Chekam2017}, C++ \cite{delgado2017}, C\# \cite{derezinska2011}, Java \cite{pitest2014}, JavaScript \cite{Mirshokraie2013}, Ruby \cite{li2015}, Android \cite{deng2017} and web applications \cite{praphamontripong2010}. It has also been adapted for some very popular programming paradigms such as Object-Oriented \cite{ma2002}, Functional \cite{le2014}, aspect-oriented and declarative programming \cite{omar2012}\cite{tuya2007}. However, to the best of our knowledge, there is no research paper that introduce mutation testing to the smart contract.

With appropriate mutation operators exposing potential defects, mutation testing can provide a strong standard for evaluating test adequacy \cite{papadakis2019mutation}. In addition to the adequacy evaluation of a test suite, mutation testing can also 
simulate the real defects of the software under test by applying mutation operators,
thus assisting the validity evaluation of the test methods proposed by researchers.
For example, Andrews et al. \cite{andrews2005} and Do et al.\cite{do2006} have proved that mutation defects generated by mutation operators are similar to real defects in effectiveness evaluation.

\subsection{Smart Contract Testing}

Chia et al. list four approaches to help blockchain test engineers in his paper \cite{Chia2018}. The first approach is to improve 
the Documentation on smart contract for developers and testers, the second approach is to fuzz the inputs of the 
smart contracts, the third approach is to mutate the code of smart contracts, and the fourth approach is to search 
the blockchain for traces of already deployed smart contracts. There are already some research results in these areas.
For example, Jiang et al.\cite{jiang2018} proposed the ContractFuzzer for the fuzzing test, which can detect vulnerabilities in smart 
contracts through random fuzzing. Wang et al.\cite{Wang2019} guides the automatic generation of efficient test cases
by tracking the execution information of smart contracts on the chain. As for mutation testing, as far as we know, 
there has been no existing literature on the application of mutation testing for smart contracts. It has been proposed
before to develop a mutation tool for smart contracts \cite{mutation-test-support}, but the attempt has been abandoned. The challenge is for 
the mutation generator to understand enough of the semantics of the smart contract language to generate only useful
mutants. 

Currently, we have found two mutation testing tools available for smart contract on GitHub, eth-mutants \cite{eth-mutants} and 
universalmutator \cite{groce2018}. However, eth-mutants only implements boundary condition mutation operators, which means that it 
can only replaces $<$ and $>$ for $<=$ and $>=$ and vice-versa. And universalmutator is a regexp based tool for mutating generic
source code across numerous languages, it has designed several mutation operators for solidity, but it’s not enough. 
Therefore, it is necessary to conduct an in-depth research to improve the effectiveness of mutation testing for smart contracts.

\section{Conclusion and Future Work}

As an effective method of improving the adequacy of testing, mutation testing is hard to be widely used in industry because of its high cost. But in the field of Ethereum, mutation testing can work well because ESC is almost impossible to be modified on Ethereum. In this paper, we proposed a novel mutation approach for ESC, together with a set of ESC specific mutation operators. The empirical study on a set of smart contracts in four real-world DApp verified that mutation testing works well in evaluating the adequacy of ESC test-suite and the proposed ESC operators can reflect real defects in practice. We believe that these mutation  operators can indeed help testers discover potential defects in smart contracts and write more adequacy test cases to ensure the security of blockchain applications.

\bibliographystyle{./IEEEtran}
\bibliography{./main}

\end{document}